\documentclass[conference,comsoc]{IEEEtran}

\IEEEoverridecommandlockouts

\usepackage[T1]{fontenc}

\usepackage{graphicx}

\usepackage{cite}
\usepackage{amsthm}
\usepackage{makecell}
\usepackage{threeparttable}
\usepackage{todonotes}
\usepackage{xcolor}
\usepackage[ruled,vlined]{algorithm2e}

\SetKwInput{KwInput}{Input}                
\SetKwInput{KwOutput}{Output}

\ifCLASSINFOpdf
 
\else
  
\fi

\usepackage{amsmath}

\usepackage[cmintegrals]{newtxmath}
\usepackage{dblfloatfix}

\usepackage{setspace}
\usepackage{fancyhdr}
\begin{document}
%

\title{Energy Efficient Data Recovery from Corrupted LoRa Frames\vspace{-6mm}}



\author{%
\IEEEauthorblockN{%
Niloofar Yazdani\IEEEauthorrefmark{1}\IEEEauthorrefmark{2},
Nikolaos Kouvelas\IEEEauthorrefmark{2},
R Venkatesha Prasad\IEEEauthorrefmark{2},
Daniel E. Lucani\IEEEauthorrefmark{1}
}%
\IEEEauthorblockA{%
\IEEEauthorrefmark{2}Embedded and Networked Systems, Delft University of Technology, the Netherlands\\
\IEEEauthorrefmark{1}DIGIT and Department of Electrical and Computer Engineering, Aarhus University, Denmark\\ \texttt{\{n.yazdani,daniel.lucani\}@ece.au.dk},
\texttt{\{n.kouvelas,r.r.venkateshaprasad\}@tudelft.nl} 
}
}

\markboth{IEEE Communications Letters}%
{Shell \MakeLowercase{\textit{et al.}}: Bare Demo of IEEEtran.cls for IEEE Communications Society Journals}

\maketitle
\thispagestyle{fancy} 
\global\csname @topnum\endcsname 0  
\global\csname @botnum\endcsname 0
\newtheorem{example}{Example}
\newtheorem{lemma}{Lemma}
\newtheorem{corollary}{Corollary}[lemma]

\vspace{-6mm}
\begin{abstract}
High frame-corruption is widely observed in Long Range Wide Area Networks (LoRaWAN) due to the coexistence with other networks in ISM bands and an Aloha-like MAC layer. LoRa's Forward Error Correction (FEC) mechanism is often insufficient to retrieve corrupted data.
%
In fact, real-life measurements show that at least one-fourth of received transmissions are corrupted.  
When more frames are dropped, LoRa nodes usually switch over to higher spreading factors (SF), thus increasing transmission times and increasing the required energy. 
This paper introduces ReDCoS, a novel coding technique at the application layer that improves recovery of corrupted LoRa frames, thus reducing the overall transmission time and energy invested by LoRa nodes by several-fold. ReDCoS utilizes lightweight coding techniques to pre-encode the transmitted data.
Therefore, the inbuilt Cyclic Redundancy Check (CRC) that follows is computed based on an already encoded data. 
At the receiver,  we use both the CRC and the coded data to recover data from  a corrupted frame beyond the built-in Error Correcting Code (ECC).  
We compare the performance of ReDCoS to (i)~the standard FEC of vanilla-LoRaWAN, and to (ii)~RS coding applied as ECC to the data of LoRaWAN. The results indicated a 54x and 13.5x improvement of decoding ratio, respectively, when 20 data symbols were sent. Furthermore, we evaluated ReDCoS on-field using LoRa SX1261 transceivers showing that it outperformed RS-coding by factor of at least 2x (and up to 6x) in terms of the decoding ratio while consuming 38.5\% less energy per correctly received transmission.
\end{abstract}

\vspace{0mm}
\begin{IEEEkeywords}
LoRaWAN, Data recovery, Energy, Error-correcting codes, CRC.
\vspace{-2mm}
\end{IEEEkeywords}

\IEEEpeerreviewmaketitle

\vspace{-1mm}
\section{introduction}
Numerous heterogeneous applications in Smart Cities and Industry 4.0 operate 
with devices that are usually energy constrained and require single-hop transmissions over large distances. 
To address the above need, the advances in RF-technology led to the rise of Low Power Wide Area Networks (LPWAN), providing energy efficient and long distance communications to IoT-devices. 
\begin{figure}
\center
\includegraphics[width=0.375\textwidth]{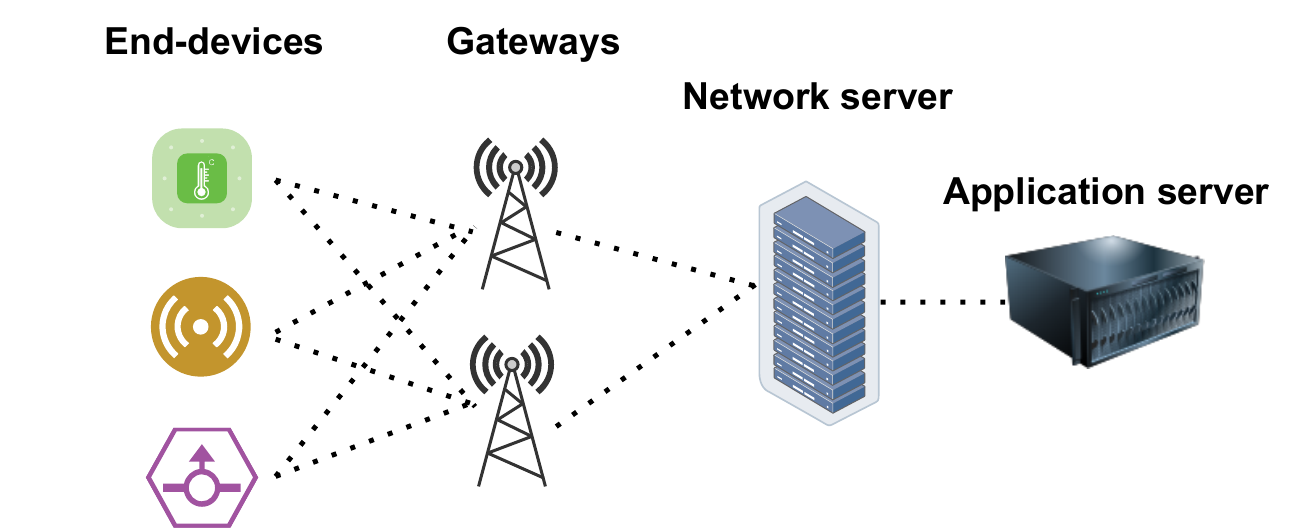}
\vspace{-5mm}
\caption{A typical LoRaWAN network}
\label{lora}
\end{figure} 
Long Range WAN (LoRaWAN)~\cite{sornin2015lorawan} is a popular LPWAN technology which --compared to others (e.g., NB-IoT, SigFox) -- offers a ``plug \& play’’ approach, providing an easily accessible network. 
LoRa networks use sub-GHz license-free ISM bands, allowing their deployment without costs for spectrum usage. 
LoRaWAN enables inexpensive, and secure transmission of small payloads over large time intervals. 
The physical layer of LoRaWAN is based on LoRa and is proprietary, owned by Semtech. 
LoRa uses Chirp Spread Spectrum~(CSS) mechanism. 
CSS enables long-range data transmissions and offers increased robustness to LoRa-signals, as they get detected even below the noise-level~\cite{Juha, Liando}. 
Fig.~\ref{lora} shows a typical LoRaWAN network, operating under a star of stars topology. As observed in Fig.~\ref{lora}, the data transmitted by the end-devices can be received by one or more gateways. 
The gateways forward the data -- using the Internet -- to the network server which is responsible for removing the duplicate messages and forwarding them to the application server. 
LoRa networks offer a wide range of configurable transmission possibilities -- Spreading Factor (SF), transmission power, carrier frequency, channel bandwidth, forward error correction, and coding rate~($CR$). 
LoRa encodes each symbol, $s$, by $N = 2^{SF}$ \textit{chips} 
and $SF \in {7, 8, ..., 12}$. Accordingly, a symbol $s$ is of length $SF$ bits and $s \in \mathcal{S}$ where $\mathcal{S} = {0, 1, ..., N-1}$~\cite{Semtech,LoRaAlliance}. 
Low $SF$s have high data rates --reaching up to 50\,kbps-- and relatively shorter transmission times but support shorter transmission ranges. 
\vspace{-2mm}
\subsection{Frame Corruption in LoRaWAN}
Since the range of LoRaWAN is very high and the power limited because of the use of unlicensed band the Received Signal Strength Indicator (RSSI) is very low, \textit{e.g.,} -115\,dBm. Invariably the LoRa transmissions often suffer from symbol corruption including bursts of symbols. This renders LoRa-frames useless, and happens due to the following: 
(i) LoRa-frames can reach gateways located a few kilometers away; around 5\,km in urban and 15\,km in rural scenarios\cite{CentenaroReview}. Over such long distances the corrupting effect of multipath propagation and physical phenomena, like shadowing and scattering, is predominant. In Non-Line of Sight (NLoS) conditions urban clutter intensifies the above phenomena. (ii) By operating in ISM bands LoRaWAN shares spectrum with other contending networks which interfere with LoRa-frames, and their transmissions corrupt LoRa-symbols.  (iii) The Medium Access Control (MAC) layer of LoRaWAN is unslotted and Aloha-like, wherein every device transmits its data upon the generation without sensing the channel for any ongoing transmission. This aggravates further the issue of corrupted frames as it increases collisions. 

In LoRaWAN, since there are no ACKs frame loss can lead to multiple issues. It leads to increased energy consumption due to  retransmissions or in this case many more redundant frames or higher sampling rate to account for lost data points. Increased traffic may lead to more collisions and more data loss resulting in spiraling of higher energy consumption. Thus the recovery of LoRa-transmissions is crucial to sustaining the battery life of the already constrained LoRa-devices as well as using the spectrum efficiently. 

For data recovery, LoRa-modems use Forward Error Correction (FEC), and specifically Hamming codes\cite{elshabrawy2019evaluation}, allowing $CR$ of $\frac{4}{4+x}$ where $x \in \{1,2,3,4\}$. Namely, every $4$~bit part of the frame is encoded into $4+x$ bits. 
$CR$-values of 4/5 and 4/6 are capable only of error detection, while 4/7 and 4/8 can correct single errors or detect double errors. 
Uplink messages transmitted by end devices can use $4$-bit Cyclic Redundancy Check (CRC) for their headers and $16$-bit CRC at the end of their payloads for error detection. 
Moreover, the payload itself ends with $32$-bit Message Integrity Code~(MIC), i.e., a Cipher-based Message Authentication Code~(CMAC) that assigns a specific signature to every end-device.
\vspace{-2mm}
\subsection{Constraints and Contributions}
As observed through real-case experiments performed by Rahmadhani and Kuipers\cite{Rahmadhani} and Marcelis \textit{et al.} \cite{Marcelis}, the correction capability of the simple FEC mechanism of LoRa gets outperformed by the sheer numbers of corrupted symbols, which often lead to failed CRC-checks and dropped frames.  Rahmadhani and Kuipers showed up to 32\% of transmissions being corrupted in experiments wherein gateways were positioned 0.5-21.5\,km away. Marcelis \textit{et al.} observed up to 53\% frame loss due to corruption when the range increases to 6\,km


Additionally, we performed real-case experiments using LoRaWAN modules to investigate the level of corruption in LoRa-transmissions. Out of the 5000 frames transmitted per case, we found 29\%-53\% being received with corrupted symbols while using the lowest transmission power. 
The experiments took place in Line of Sight (LoS), with an operating frequency of $868.1$~MHz, a bandwidth of $125$~kHz on SF8 and SNR of around 11.50. 

It is easy to see from the Table.~\ref{ta:rec} -- which  depicts the portion of corrupted frame-payloads in our experiments -- that it is necessary to circumvent the frame corruption. Thus the question  we  try  to  answer  here  is: \textbf{Can  we  recover  the data  even  if  CRC  check  fails?}

\begin{table}
\begin{center}
\begin{threeparttable}[b]
	\footnotesize
	\caption{Frame Reception Measurement}
	\label{ta:rec}
	\begin{tabular}{|c|c|c|}
		\hline
		payload size~[B]  & uncorrupted Received & Corrupted Received\\
		\hline
	    $14$ & $35.24\%$ & $28.45\%$\\
		\hline
		$18$ & $45.28\%$ & $36.71\%$\\
		\hline
		$24$ & $16.54\%$ & $53.25\%$\\
		\hline
		$28$ & $23.93\%$ & $49.72\%$ \\
		\hline 
	\end{tabular}
\end{threeparttable}
\end{center}
\end{table}
However, any frame recovery algorithm needs to adhere to the constraints posed by LoRaWAN: 
(i) No changes should be made to the gateways and the existing infrastructure of LoRa networks. 
(ii) No ACK from the gateway is expected since it increases the message overhead and the energy consumption due to extra listening times. 
(iii) The introduced symbol redundancy due to encoding should be minimal, especially for high SFs, because the allowed payload size is limited (i.e., 51\,B for SF10-SF12).

To this end, we introduce ReDCoS, \underline{Re}covery of \underline{D}ata of \underline{Co}rrupted \underline{S}ignals; a novel coding scheme at the application layer of LoRaWAN. ReDCoS operates proactively before the LoRa-FEC mechanism is applied. 
%
Our contributions are the following:
\begin{enumerate}
    \item We design ReDCoS, a novel application layer coding technique that allows the recovery of LoRa transmissions well beyond built-in error-correcting capability (see Sect.~\ref{system}). 
    \item We provide a mathematical analysis of ReDCoS, finding the probability of successful decoding (see Sect.~\ref{analysis}). 
    \item We evaluate the performance of ReDCoS not only numerically but also experimentally, by conducting on-field experiments using LoRa-devices and a gateway (see Sect.~\ref{evaluation}). 
\end{enumerate}
 

\vspace*{-5mm}
\section{Related Works}
Marcelis \textit{et al.} introduced DaRe, a coding scheme against data-loss, operating at the application layer of LoRaWAN. DaRe combines convolutional and fountain codes, providing 21\% higher data recovery than LoRaWAN's repetition coding\cite{Marcelis}. 
Coutaud and Tourancheau designed CCARR which uses Reed Solomon (RS) at the frame-level. CCARR transmits encoded super-frames, involving application data and redundancy data (RS-frames). 
CCARR guarantees high packet delivery ratio (even 100\%) over lossy channels\cite{Coutaud}. 
Borkotoky \textit{et al.}  propose two application-layer coding schemes --windowed and selective coding-- for delay-intolerant LoRaWANs with minimum feedback. Windowed mechanism encodes by accounting all the non-delivered and non-expired transmitted symbols, while selective coding picks certain among them based on feedback\cite{Borkotoky}. 
Tsimbalo \textit{at al.} exploit the redundancy that CRC adds to the encoded data by applying iterative algorithms -- ADMM and BP -- on the decoding side. They correct up to 35\% of corrupted BLE-packets without consuming any extra energy at the transmitter side\cite{Tsimbalo}. 
Sant'Ana \textit{et al.} combine two coding techniques of LPWANs: transmission of packet-replicas, RT, and transmission of encoded packets using linear operations like XORs, CT. Their hybrid coding scheme, which is optimized for LoRaWAN, can support the transmissions by more devices than RT and CT alone while extending their battery lifetime\cite{SantAna}. 
Chen \textit{et al.} extend the lifetime and communication range of LoRaWAN, by introducing eLoRa, which recovers the correct parts of a packet by applying Luby codes (LT) on several versions of the packet that are received by multiple gateways. Further, eLoRa optimizes physical layer parameters to best serve lifetime and range requirements\cite{Chen}. 
Coutaud \textit{et al.} characterize the behavior of LoRa-channels in urban scenarios and adapt the specifics of ADR's FEC and retransmission schemes, in order to improve the current Adaptive Data Rate (ADR) mechanism of LoRaAWAN. \cite{CoutaudADR}. 

\vspace{-5mm}
\section{Background}

Cyclic Redundancy Check~(CRC) are error-detecting codes used by communication and storage systems to detect corruption of data for instance caused by noise. To encode the data, considering the systematic approach, CRC-$m$ appends a check value of a fixed length of $m$ bits to the data using a hash function to map the data values to the $m$ bits.  Upon retrieval, if the data is corrupted the calculated CRC value does not match the appended value, thus a CRC error occurs. 

Reed-Solomon codes (RS)~\cite{wicker1999reed} are a family of linear Error-Correcting Codes~(ECC) used to correct errors in many applications including communication systems and storage systems.
RS codes treat a data block, i.e., \textit{message}, as a series of symbols from a finite field of size $q$, $\mathbb{F}_q$. A RS code is specified as $RS_q(n,k)$ where $q$ denotes the finite field whose symbols are used, and $k$, $n$ are the symbol-size of the message and the codeword, respectively, where $k < n$.  According to the above, upon the arrival of a message of length $k$ symbols, $\{x_1, x_2,..., x_k\}$ where $0 \le x_i < q$, the encoder generates a codeword of length $n$ symbols, $\{y_1, y_2,..., y_n\}$ where $0 \le y_i < q$. The codeword length, $n$, is equal to $q-1$; However, using a short version of RS codes~\cite{moreira2006essentials}, $n$ is more flexible, \textit{i.e.}, $k < n < q$. 
Let $t = n - k$, a Reed-Solomon code can correct up to and including $\lfloor \frac{t}{2} \rfloor$ corrupted symbols at unknown locations.
An RS code can also be used as an \textit{erasure code} where it can correct up to and including $t$ erasures at \textbf{known locations}. In this paper, we use the functionality of RS codes as an erasure code.
The encoding procedure of RS codes can be \textit{systematic}, wherein messages appear at the beginning of the codeword followed by $t$ error correcting symbols. 

\section{Proposed Scheme: Data Recovery Beyond Built-In Error-Correction}\label{system}
\begin{figure}[t]
\center
\includegraphics[width=0.4\textwidth]{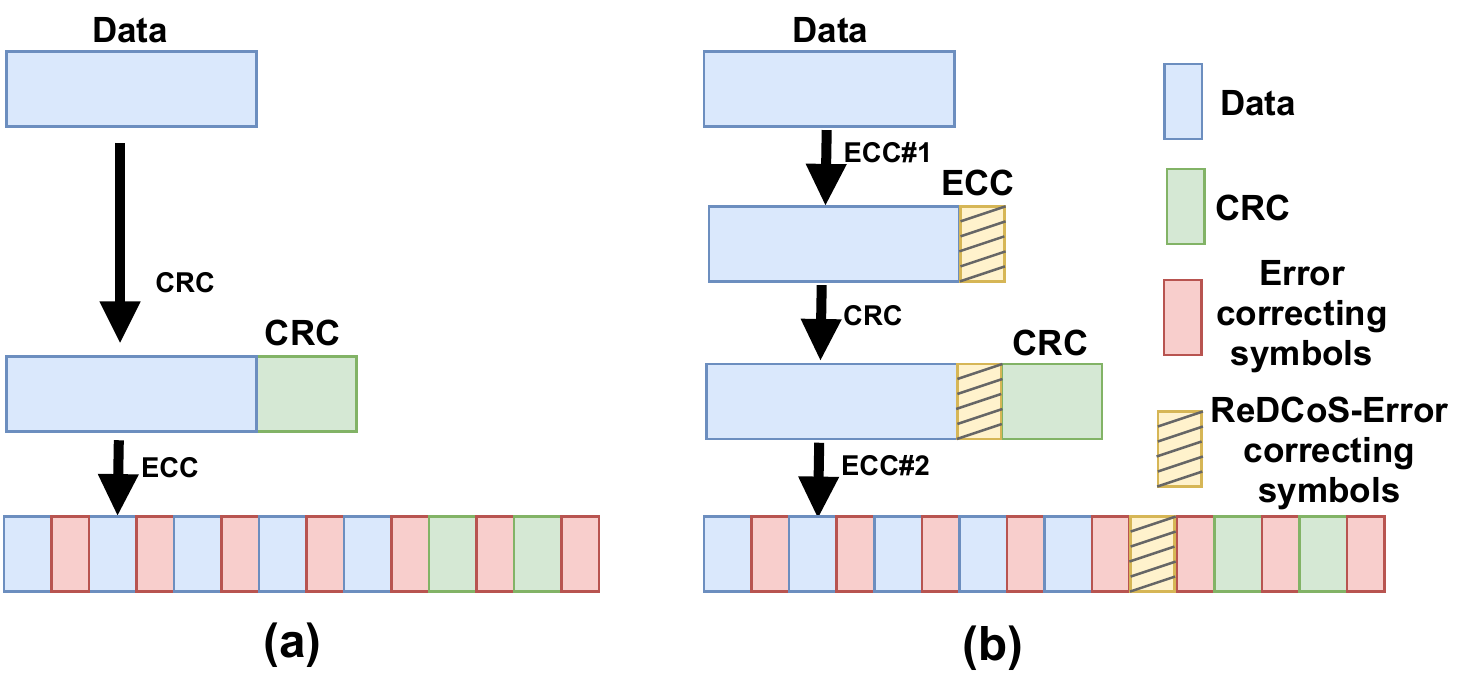}
\vspace*{-4mm}
\caption{The order of using a CRC and an ECC (a) Conventional method, (b) proposed scheme: ReDCoS.}
\label{baseidea}
\end{figure}

In a typical wireless system, \textit{e.g., WiFi}, the sender typically takes the data, computes the CRC, and applies an ECC, \textit{e.g.}, Reed-Solomon or Hamming, on the data plus the CRC as shown in Fig.~\ref{baseidea}.~(a).  Note that in this paper, we consider using a systematic encoding procedure for the sake of simplicity. 
On retrieval, the receiver applies the decoding function of the ECC to retrieve the data and the CRC and to correct possible corrupted bits/symbols. Afterwards, the receiver calculates the CRC for the recovered data.  If the calculated CRC matches the received CRC, data is received successfully. Otherwise, data cannot be retrieved and this happens if the number of corrupted symbols exceeds the correction capability of the applied ECC. We try to recover the data even if CRC check fails. 
%
\vspace{-1mm}

\textbf{The Idea.} The idea is to make the system more robust by adding a limited number of extra symbols using an ECC as shown in Fig.~\ref{baseidea}.~(b). 
This redundancy is used later to create the correct CRC and the correct data using majority voting. 
The recovered CRC is later verified using parts of the received CRC that are not damaged. The important technique here is to \textbf{combine the concepts of the CRC and ECCs to utilize ECCs beyond their error-correcting capability.}

\textbf{Encoder.} 
Encoder is lightweight and suitable for resource limited devices, as it only appends a few extra symbols to the end of the data using the encoding function of an ECC~(ECC\#1, referred to as ReDCoS' ECC, hereafter). Such ECCs can be RS codes, as shown with yellow color (dashed area) in Fig.~\ref{baseidea}.~(b).
The rest of the process is similar to a typical wireless system where a CRC is calculated for the encoded data, and the encoding function of an ECC~(ECC\#2) is applied to the encoded data plus the CRC as shown in Fig.~\ref{baseidea}.~(b). The two ECCs are applied independently.
Note that by \textit{encoded data}, hereafter, we refer to the data plus the few extra symbols added by the ReDCoS' ECC. In this paper, we consider CRC-$32$.
We define $l$ as the symbol-size of the CRC.
\vspace{-4mm}
\begin{algorithm}
	\scriptsize
	\DontPrintSemicolon
	\caption{ReDCoS' decoder:}
	\KwInput{$da\_r$: received data, $ecs\_r$: received error correcting symbols, $CRC\_r$: received CRC, $H$: minimum number of CRC symbol matches}
    \KwOutput{$da\_u$: uncorrupted data, failed decoding}
    $CRC\_t \leftarrow \{\}$ \tcp*{\scriptsize{pairs of (CRC:\#repetition)}}
	\texttt{Step 1:} Check if the received encoded data is uncorrupted.\;
	$i \leftarrow find\_CRC(da\_r, ecs\_r)$\;
	\If{$i == CRC\_r$}{ 
	    $return(da\_u = da\_r)$ \tcp*{\scriptsize{End: SUCCESS}}
	}
	\For{$j = \{1,2,...,C(k+t,k)\}$}{
	    \texttt{Step 2:} Generate an uncorrupted encoded data candidate: Try a $k$-combination\;
	    $y \leftarrow \{\}.size(k+t)$\;
	    $y \leftarrow select\_k\_symbols(da\_r,ecs\_r)$\;
	    $y \leftarrow calculate\_remaining\_t\_symbols(y)$\;
	    $i \leftarrow find\_CRC(y)$\;
	    \texttt{Step 3:} Check if the candidate matches the received CRC\;
	    \If{$i == CRC\_r$}{ 
	    $da\_u = pick\_first\_k\_symbols(y)$\; 
	    $return(da\_u)$ \tcp*{\scriptsize{End: SUCCESS}}
	    }
	    \texttt{Step 4:} Save/Check the recovered CRC.\;
	    \tcp*{\scriptsize{Check key $i$ availability in CRC table $CRC\_t$}}
	    \If{$!find\_key(i,CRC\_t)$}{
	    $CRC\_t[i] = 1$\;
	    }
	    \Else{
	    $CRC\_t[i] += 1$\;
	    \texttt{Step 5:} Majority voting.\;
	    \If{$CRC\_t[i] == k+1$}{
	    $x \leftarrow find\_\#similar\_symbols(i,CRC\_r)$\;
	    \texttt{Step 6:} Verify the recovered CRC.\;
	    \If{$x \geq H$}{
	    $da\_u = pick\_first\_k\_symbols(y)$\;
	    $return(da\_u)$ \tcp*{\scriptsize{End: SUCCESS}}
	    }
	    }
	    }
	}
	\texttt{Step 7:} return failure if algorithm has not ended yet by a return\;
	$return("Failed\; Decoding")$  \tcp*{End: Failed}\;
	
	\label{alg:decoder}
	\vspace{-4mm}
\end{algorithm}
\vspace{-2mm}
\begin{figure*}
\center
\includegraphics[width=0.8\textwidth]{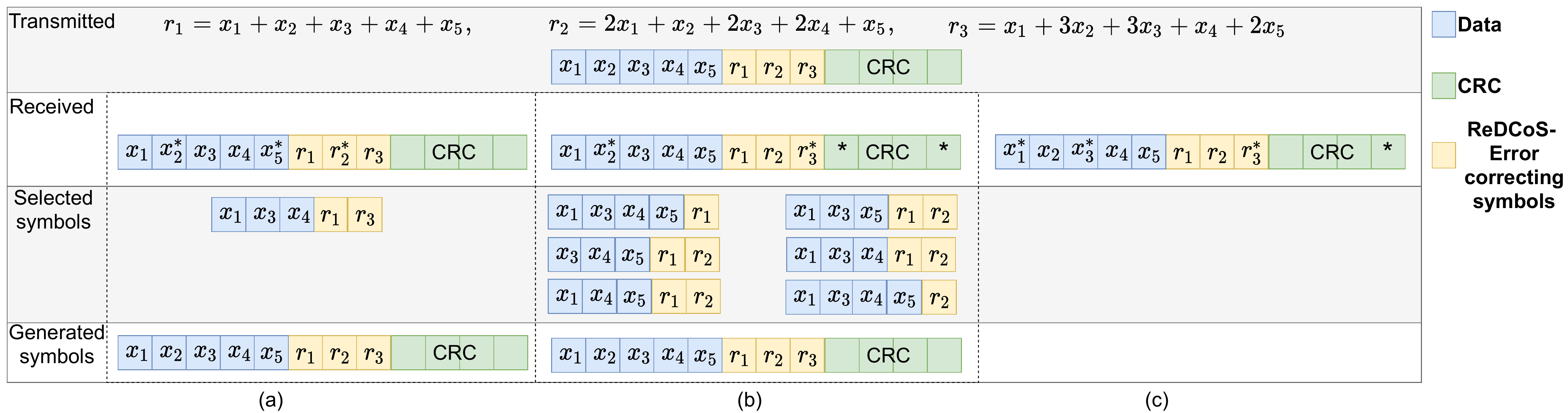}
\vspace{-4mm}
\caption{Data retrieval examples, a) Success using the received uncorrupted CRC, b) Success using the partially uncorrupted received  CRC. c) Failure.}
\vspace{-4mm}
\label{idea}
\end{figure*} 

\textbf{Decoder.} When a corrupted frame is received the decoding of second ECC is done to recover the encoded data plus the CRC. 
The formal decoding algorithm for a given received payload is detailed in
algorithm~\ref{alg:decoder}, where $k$ denotes the symbol size of the data and $t$ denotes the symbol size of the error correcting symbols added by ReDCoS' ECC.
We also use $C(z,p)$ to represent the number of $p$-combinations from a given set of $z$ elements where $C(z,p) = \frac{z!}{p!(z-p)!}$.
The parameter $H$ denotes the minimum number of required symbols matching the recovered CRC and the received CRC to verify the recovered CRC.
According to algorithm~\ref{alg:decoder}, ReDCoS recovers the data if one of the following cases happens:
(i) The received encoded data and the received CRC are both uncorrupted.
(ii) The received CRC is uncorrupted and there is a minimum of $k$ correct symbols out of the $k+t$ symbols of the received encoded farmes. 
(iii) A minimum of $H$ symbols of the received CRC are uncorrupted~(to verify the recovered CRC) and a minimum of $k+1$ symbols of the received encoded data are uncorrupted.
In the last case, the recovered uncorrupted CRC will be repeated a minimum of $k+1~(C(k+1, k))$ times by trying different $k$-combinations from the given $k+t$ symbols for the encoded data. There are a total of $C(k+t,k)$ $k$-combinations that a decoder can try in brute force.
\textit{Majority voting}, in this paper, denotes finding CRCs which are repeated a minimum of $k+1$~times.  Selecting the $k$-combinations could happen at random; however, a given $k$-combination should not be tried two times.
By selecting a $k$-combination, the decoder calculates the remaining $t$ symbols (that were not picked randomly by the decoder) for the uncorrupted encoded data candidate using the decoding and encoding functions of the ReDCoS' ECC.
In this paper, particularly, we concentrate on: (i)~the encoded data plus the CRC, \textit{i.e.,} before applying the encoding function of the second ECC at the sender side, and (ii)~the received encoded data plus the CRC, \textit{i.e.,} after applying the decoding function of the second ECC at the receiver side.

%
\begin{figure*}
\center
\includegraphics[width=0.8\textwidth]{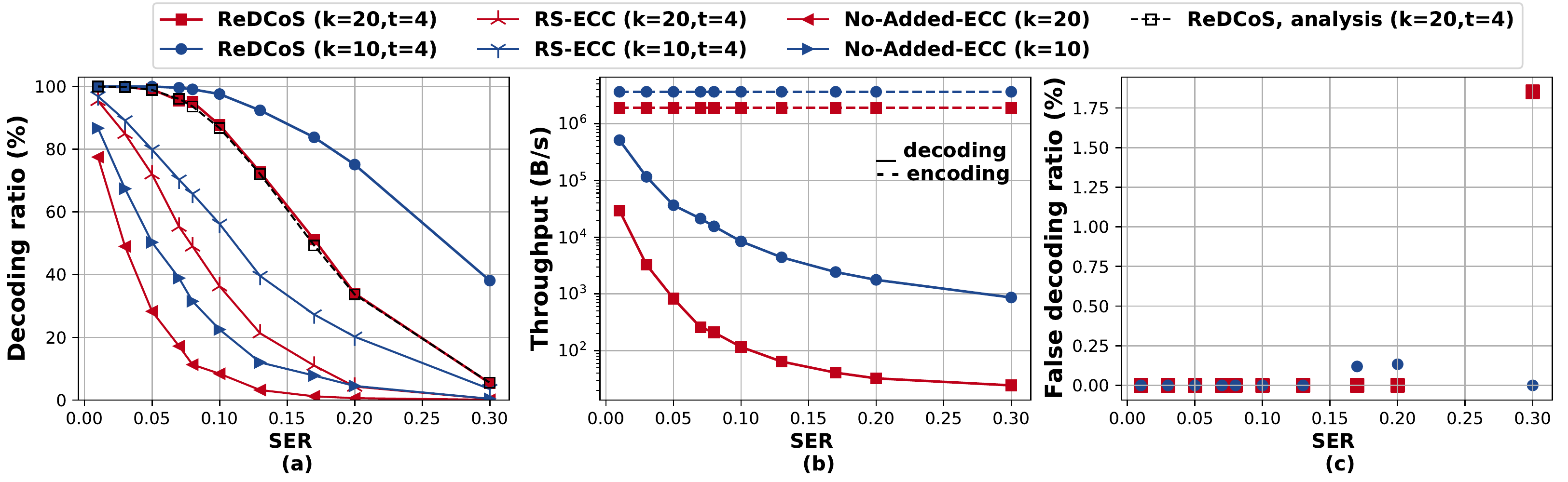}
\vspace{-4mm}
\caption{(a) Decoding ratio, (b) encoder's and decoder's throughput, (c) false decoding ratio, all versus SER.}
\vspace{-4mm}
\label{sim}
\end{figure*}
\vspace{-2mm}
\begin{example} \label{ex:scheme}
Fig.~\ref{idea} illustrates $3$ examples for ReDCoS' decoder, including $2$ successful decoding and $1$ failed decoding.
Let us take the data with $5$ symbols $\{x_1, x_2, x_3, x_4, x_5\}$, \textit{i.e.,} $k = 5$.
The data is encoded as $\{x_1, x_2, x_3, x_4, x_5, r_1, r_2, r_3\}$ where $t=3$ and $r_1, r_2$, and $r_3$ are independent linear combinations of $x_1, x_2, x_3, x_4$, and $x_5$ (the operations are over a given finite field).
We assume that some of the symbols are corrupted while receiving.
The corrupted symbols are shown with a star. The number of initially corrupted symbols could be higher but some are already corrected by the second ECC. 
\end{example}
In particular, we consider $3$ scenarios:

\textbf{Example\ref{ex:scheme}.1.} 
$5$ symbols out of $k+t$ symbols are uncorrupted while $5=k$. 
While decoder is trying different $k$-combinations, as soon as it selects the error-free symbols, \textit{i.e.,} $x_1, x_3, x_4, r_1, r_3$, it will construct the correct encoded data and hence it generates the correct CRC.  
As the received CRC is not corrupted, the generated CRC would be identical to the received CRC and data is retrieved successfully. 

\textbf{Example\ref{ex:scheme}.2.} $2$ symbols out of $k+t = 8$ symbols are corrupted. 
Furthermore, $2$~symbols of the CRC are corrupted as shown in Fig.~\ref{idea}.~(b). 
This means that $6$ symbols out of $k+t$ symbols are uncorrupted where $6 = k+1$ and $2$ symbols of the CRC are uncorrupted as well (considering a $4$-symbol CRC).
The decoder tries different $k$-combinations which will lead to different CRC values.
A total of $C(8,5)=56$ different combinations exist. 
However, every time the receiver selects $k = 5$~symbols out of the $6$ correct symbols, \textit{i.e.}, $x_1, x_3, x_4, x_5, r_1, r_2$, the correct data and hence the correct CRC will be generated. 
This means that the correct CRC will be repeated $C(6,5) = 6~( k+1)$~times with choosing the combinations shown in Fig.~\ref{idea}.~(b).
Verifying the recovered CRC, 2 symbols of the recovered CRC match the received CRC. 
Considering $H=2$, the recovered CRC is verified and data can be retrieved successfully.

\textbf{Example\ref{ex:scheme}.3.} $3$~symbols out of $k+t=8$~symbols are corrupted in addition to $1$ symbol of the CRC.
This means that $5$~symbols out of $k+t=8$~symbols are uncorrupted, where $5 = k$. 
The only combination which generates the correct CRC is $x_2, x_4, x_5, r_1, r_2$.
Thus, the correct CRC will be repeated only once similar to many other CRCs. 
Since the received CRC is also corrupted, there is no reliable way to retrieve the data.

\begin{figure}
\center
\includegraphics[width=0.35\textwidth]{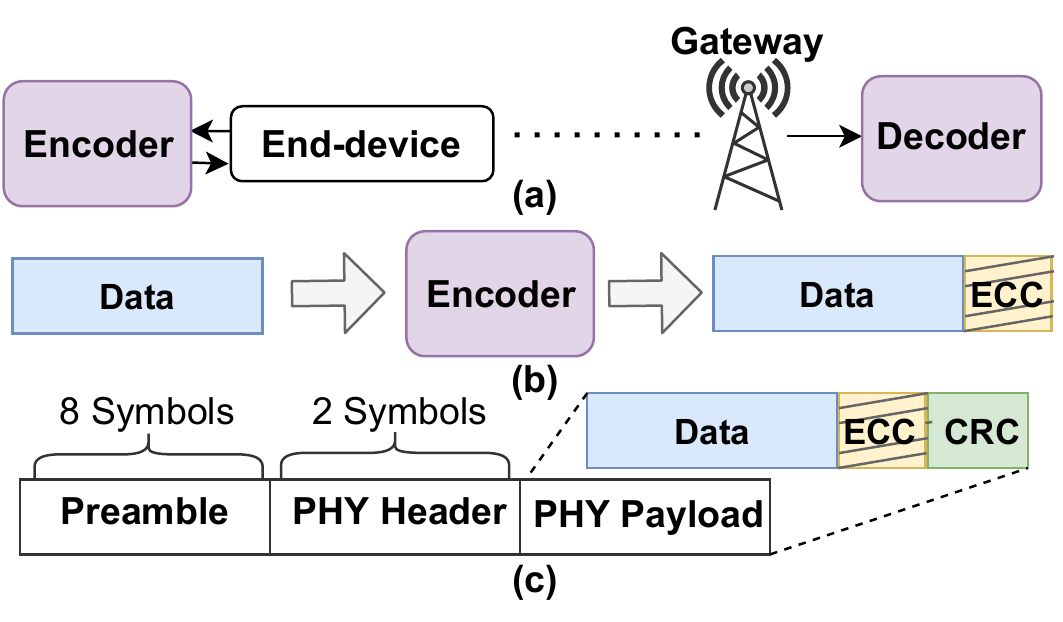}
\vspace{-4mm}
\caption{Implementation. (a) LoRa device and a Gateway. (b) The encoder is implemented as a C++ library to compute error correcting symbols. (c) LoRaWAN frame. Note that there is no change in the LoRa frame format.}
\label{implementation}
\end{figure}
\section{Mathematical Analysis of ReDCoS}\label{analysis}
\vspace{-2mm}
We evaluate ReDCoS in terms of \textbf{encoder and decoder throughput, decoding ratio, and false decoding ratio} for different Symbol Error Rates~(SER). 
SER is the ratio of corrupted symbols over transmitted symbols, and it denotes the probability of a symbol getting corrupted during the transmission.
The encoder and decoder \textit{throughput} indicate the amount of data~(i.e., excluding error-correcting symbols or CRC) processed. 
We define the \textit{Decoding Ratio~($DR$)} as the ratio of decoded over total received payloads, 
and the \textit{False Decoding Ratio~($FDR$)} as the ratio of falsely decoded over total decoded payloads. 
False decoding occurs if the recovered payload is not identical to the associated transmitted payload. 
Increasing $H$ reduces the FDR as the recovered CRC must be validated more strictly, i.e., requiring more symbol-matches.

\textbf{Decoding Probability.}
Without loss of generality, each symbol can get corrupted during transmission independently of other symbols under probability $SER$. 
$P(w,e)$ is the probability to have $e$ corrupted out of $w$ given symbols,
\vspace{-2mm}
\begin{equation}
\vspace{-2mm}
P(w,e) = SER^e \cdot (1 - SER)^{w-e} \cdot C(w,e), \forall e \in [0,w]
\end{equation} 

\textit{Received CRC is uncorrupted.}
The probability of receiving an uncorrupted CRC is $P(l,0)$.
In this case, ReDCoS recovers the data if up to and including $t$ symbols are corrupted out of the $k+t$ symbols. 
This happens with probability $\Sigma_{i=0}^{t}P(k+t,i)$.

\textit{Matched symbols between recovered and received CRC greater or equal to $H$ and less than $l$.}
The number of corrupted symbols of the CRC is  within $\{1,2,..., l-H\}$.
This happens with probability $\Sigma_{i=1}^{l-H}P(l,i)$.
In this case, ReDCoS recovers the data if up to and including $t-1$ symbols are corrupted out of the $k+t$ symbols~(equivalently, $k+1$ symbols are intact). 
This happens with probability $\Sigma_{i=0}^{t-1}P(k+t,i)$. 

\textit{Matched symbols between recovered and received CRC less than $H$.}
In this case, ReDCoS cannot decode the data.  Based on the specific probabilities of each case above, the probability of successfully decoding a frame is, 
\vspace{-2mm}
\begin{equation}
\mathbb{P}_D = P(l,0) \cdot \Sigma_{i=0}^{t}P(k+t,i) + 
\Sigma_{i=1}^{l-H}P(l,i) \cdot \Sigma_{i=0}^{t-1}P(k+t,i)
\end{equation}

\vspace{-5mm}
\section{Performance Evaluation}\label{evaluation}
\vspace{-1mm}
\begin{table*}
\begin{center}
\begin{threeparttable}
	\footnotesize
	\caption{Implementation results and characteristics}
	\label{ta:imple}
	\begin{tabular}{|c|c|c|c|c|c|c|c|}
		\hline
		\makecell{(k, t)}  & SNR(min, max, ave) & RSSI(min, max, ave)[dBm] & \makecell{Received\\ frames}\tnote{1} & \makecell{Corrupted\\ frames}\tnote{2} & \makecell{encoder\\throughput[MB/s]} & \makecell{decoder\\ throughput} & \makecell{encoder\\energy~[nJ/b]}\\
		\hline
		($10,4$) & ($-14,-9,-11.9$) & ($-116,-114,-115.1$) & $63.7\%$ & $44.67\%$ & $0.5724$  & $2087.7$~B/s &  $38.6$\\
		\hline
		($10,8$) & ($-14,-1,-11.18$) & ($-116,-113,-115.0$) & $82.0\%$ & $44.77\%$ & $0.5348$ & $128.6$~B/s & $41.3$\\
		\hline
		($20,4$) & $(-15,-2,-11.78)$ & ($-116,-112,-115.1$) & $69.8\%$ & $76.3\%$ & $0.1580$ & $39.3$~B/s & $140.0$\\
		\hline
	\end{tabular}
	$^1$ Uncorrupted and corrupted received frames. \quad
	$^2$ \# of corrupted received frames / \# of received frames. 
\end{threeparttable}
\end{center}
\vspace{-9mm}
\end{table*}
We consider $RS_{256}(k+t,k)$ as the ReDCoS' ECC for evaluation. Moreover, all LoRaWAN frames are transmitted at SF8. Thus, each symbol is $1$ byte. 
We compare the decoding ratio of the ReDCoS to two other approaches: 
(i)~using Reed-Solomon codes as an ECC~(RS-ECC) with the same number of error correcting symbols and data symbols, which is capable of data recovery if up to and including $\lfloor \frac{t}{2} \rfloor$ symbols of encoded data are corrupted and CRC is uncorrupted, and 
(ii)~without using any added ECC~(No-Added-ECC) with the same number of data symbols, wherein data recovery is possible if all $k$ data symbols and the received CRC are uncorrupted. This corresponds to vanilla-LoRaWAN.
\vspace{-2mm}
\subsection{Numerical Results}
\vspace{-2mm}
We use a Core i7-7820HQ at 2.90GHz for the tests and focus on a single thread (and CPU) for execution.
We consider $H = 2$. Fig.~\ref{sim} shows the ReDCos' performance for $2$ different cases: ($k=10, t=4$), and ($k=20,t=4$) over $10$ different values for SER. 
The measurements average over $1000$ runs per case per SER. 
It is assumed that all the transmitted frames are received by the receiver.
Fig.~\ref{sim}.~(a) depicts the decoding ratio to SER.
In general, as the values of SER increase the decoding ratio decreases for all $3$ schemes. 
As shown in Fig.~\ref{sim}.~(a), for ($k=20,t=4$), ReDCoS outperforms RS-ECC and No-Added-ECC in terms of decoding ratio by up to $13.5$ and $54.0$ times, respectively. 
Moreover, for ($k=10,t=4$), ReDCoS outperforms by up to $10.8$ and $95.2$ times, correspondingly. 
%
As depicted in Fig.~\ref{sim}.~(b), ReDCoS' encoding throughput is $1.90$~MBps and $3.65$~MBps for ReDCoS($k=20,t=4$) and ReDCoS($k=10,t=4$), respectively.
As shown in Fig.~\ref{sim}.~(b), ReDCoS' decoding throughput decreases with the increase of SER because the number of corrupted symbols increases.
The decoding throughput for ReDCoS($k=20,t=4$) ranges from $29.6$~kBps for SER=$0.01$ to $24.4$~Bps for SER=$0.3$.
The decoding throughput for ReDCoS($k=10,t=4$) ranges from $513.7$~kBps for SER=$0.01$ to $ 865.9$~Bps for SER=$0.3$.
Fig.~\ref{sim}.~(c) shows FDR regarding ten different SER values. For ReDCoS ($k=20,t=4$) and ($k=10,t=4$), $FDR=0$ for nine and eight out of ten cases, respectively. 
SER value can be decreased by increasing $H$.
\vspace{-2mm}
\subsection{Practical Evaluation}
\vspace{-2mm}
For our experiments, we consider one end-device~(sender) and one gateway~(receiver) as shown in Fig.~\ref{implementation}.~(a). 
We used LoRa SX1261 nodes, integrated with a Raspberry Pi 3 Model B.
The sender and receiver antennas are isotropic with $0$\,dBi gain. 
The operating frequency and bandwidth are $868.1$~MHz and $125$~kHz, respectively. 
The encoder is implemented as a C++ library that is included in the end-device software.  The encoder operates on each data and generates the encoded data as shown in Fig.~\ref{implementation}.~(b).
Data contain random symbols. Fig.~\ref{implementation}.~(c) shows the transmitted LoRaWAN frame structure. 
We consider only the mandatory parts of LoRa-frames to keep the frame structure as simple as possible. 
The optional frame integrity checks and the code rate are deactivated to facilitate the reception of more corrupted frames during our experiments.
Alternatively, a $4$-byte CRC check associated with the encoded data is used as shown in Fig.~\ref{implementation}.~(c). 

\textbf{General.} We evaluated $3$ cases as shown in Table.~\ref{ta:imple}: ($k=10$, $t = 4$), ($k=10$, $t = 8$), and ($k=20$, $t = 4$).  
For each case, $5000$ frames are transmitted.
As depicted in Table.~\ref{ta:imple}, the SNR~(Signal-to-Noise Ratio) value and the RSSI~(Received Signal Strength Indicator) value of the received frames lies between -15 to -1 and between -116dBm to -112dBm, respectively.
Best-case scenario, $82\%$ of the transmitted frames are received while this value is $63.7\%$ for ($k=10$, $t = 4$).
Up to $76.3\%$ of the received frames are corrupted. 
The decoder throughput is $2087.7$~B/s for ($k=10$, $t = 4$). Increasing $k$ or $t$ increases the average number of possible iterations for finding the correct data and hence decreases the throughput. 
As depicted in Table.~\ref{ta:imple}, the throughput for ($k=20$, $t = 4$) is as low as $39.3$~B/s.
Contrarily, the encoder is lightweight. For instance, the encoder's throughput and energy consumption are $0.5724$~MB/s and $38.6$~nJ/s, respectively for ($k=10, t = 4$).

\begin{figure}
\center
\includegraphics[width=0.33\textwidth]{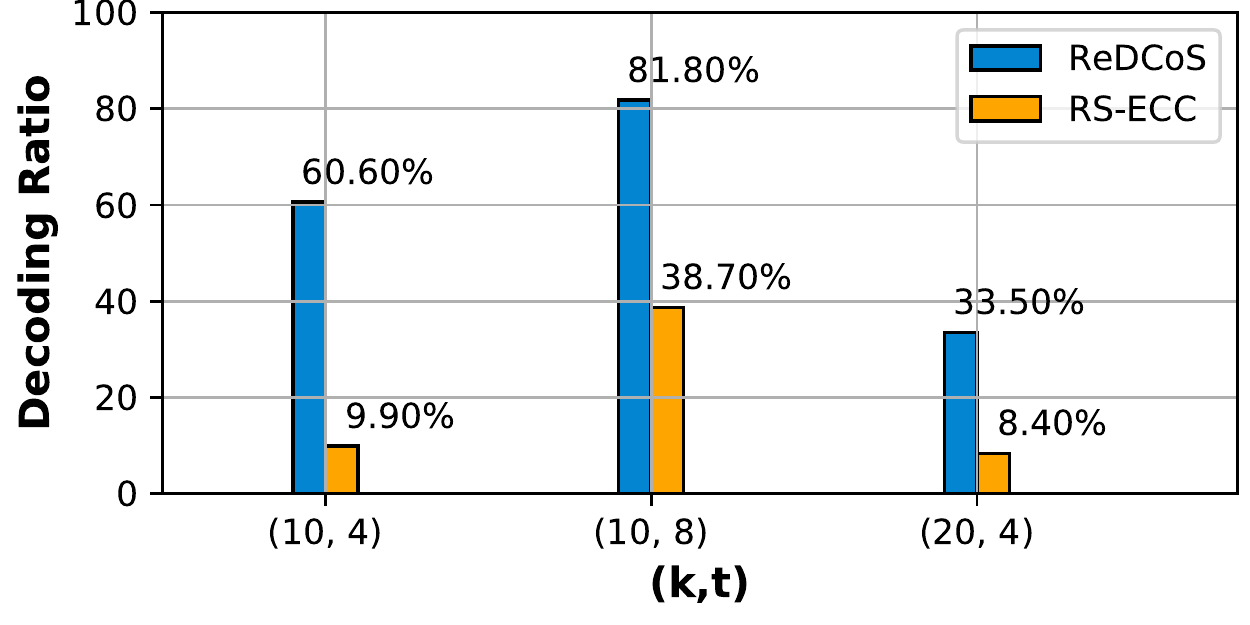}
\vspace{-5mm}
\caption{Decoding ratio considering only corrupted received frames.}
\vspace{-1mm}
\label{de-impl}
\end{figure} 
%
\textbf{Decoding and False Decoding Ratios.}
Fig.~\ref{de-impl} depicts the decoding ratio for ReDCoS and RS-ECC. 
This decoding ratio is calculated by considering only received corrupted frames, \textit{i.e.,} the decoding ratio shows the percentage of the received corrupted frames that are recovered successfully. 
As depicted in Fig.~\ref{de-impl}, ReDCoS outperforms RS-ECC in terms of decoding ratio by factors of $6.12$, $2.11$, and $3.99$ times for ($k=10$, $t = 4$), ($k=10$, $t = 8$), and ($k=20$, $t = 4$), respectively. 
Table.~\ref{ta:brief}, shows the portion of the received frames~(corrupted and uncorrupted), and the portion of correctly received frames by the three approaches as well as the ReDCoS' FDR.
As shown in Table.~\ref{ta:brief}, $FDR$ is as low as $0\%$, $0.13\%$, $0.11\%$ for the same above cases, respectively. \\
\begin{figure}
\center
\includegraphics[width=0.35\textwidth]{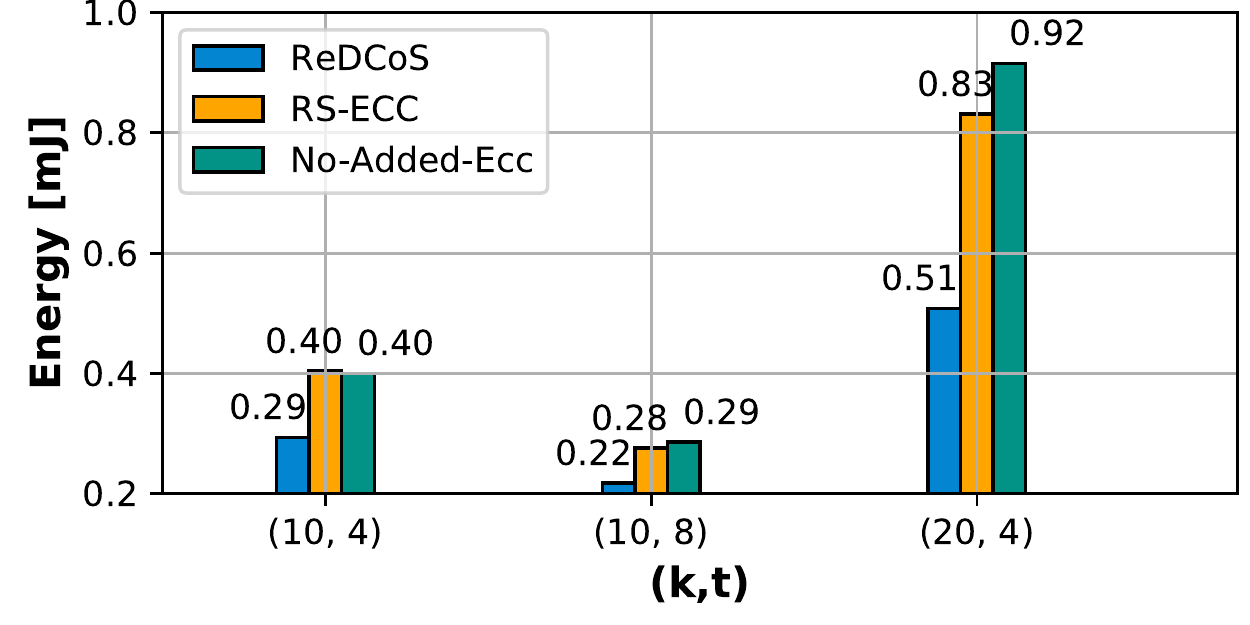}
\vspace{-4mm}
\caption{Transmission energy for one correctly received frame.}
\vspace{-2mm}
\label{energy}
\end{figure} 
\textbf{Energy:}
The main source of energy consumption in an end-device is data transmission,  
which can be calculated by multiplying the transmission power ($0$~dBm or $0.001$~W) by the frame Time on Air (ToA). ToA depends on the SF, and bandwidth used\cite{Semtech}. 
However, if a frame is not correctly received, it needs to be retransmitted or it had to be coded over multiple frames\cite{Marcelis}. To consider the effect of frame loss, the energy consumption per correctly received frame has been considered~(see Table.~\ref{ta:brief}) and is shown in Fig.~\ref{energy}. 
As seen, ReDCoS consumes up to $38.5\%$ and  $44.5\%$ less energy compared to ECC-RS and No-Added-ECC, respectively. 
\begin{table}
\begin{center}
\begin{threeparttable}
	\footnotesize
	\caption{Measurements}
	\vspace*{-1mm}
	\label{ta:brief}
	\begin{tabular}{|c|c|c|c|c|c|}
	\hline
		\makecell{(k, t)}&\makecell{Received\\frames}&\makecell{Correctly\\received,\\No-Added\\-ECC} & \makecell{Correctly\\received,\\RS-ECC} & \makecell{Correctly\\received,\\ReDCoS}&$FDR$\\
		\hline
		($10,4$) & $63.7\%$ & $35.94\%$ & $38.08\%$ & $52.48\%$ &  $0\%$\\
		\hline
		($10,8$) & $82.0\%$ & $50.36\%$ & $59.49\%$ & $75.31\%$ &  $0.13\%$\\
		\hline
		($20,4$) & $69.8\%$ & $17.96\%$ & $21.00\%$ & $34.37\%$ &  $0.11\%$\\
		\hline
	\end{tabular}
	All items are calculated considering all the transmitted frames.
\end{threeparttable}
\end{center}
\vspace{-3mm}
\end{table} 

\vspace{-2mm}
\section{Conclusions}
\vspace{-2mm}
LoRa offers a low bitrate transmission of IoT data. However, the MAC layer offers no guarantee and corrupted frames are dropped. In this paper, we proposed \underline{Re}covery of \underline{D}ata of \underline{Co}rrupted \underline{S}ignals for the application layer of LoRa networks, wherein LoRa-payloads are pre-encoded using lightweight coding scheme before the addition of the CRC. This leads to more robust encoded packets after applying LoRa-FEC. 

We provided the probabilistic analysis of successful decoding using ReDCoS. 
Further, we conducted numerical simulation and on-field evaluation of ReDCoS. 
The results show that ReDCoS outperforms manifolds not only the built-in error-correction scheme of LoRaWAN, but also RS codes when used as ECC. 
Additionally, because of ReDCoS' data-recovery capability, less packet retransmissions (or encoding over multiple frames) occur, reducing the energy consumption by 44.5\% as well as age of information while doubling the frame reception. 
ReDCoS acts independently of the standard LoRa-encoding requiring no change to infrastructure and/or protocol. Moreover, it benefits from a simple encoder.
Further, ReDCoS does not need buffer and/or feedback downlinks for its operation, keeping message-overhead low. 

\ifCLASSOPTIONcaptionsoff
  \newpage
\fi
\bibliographystyle{IEEEtran}
\vspace{-3mm}
\bibliography{ref}

\end{document}